**Stakeholder perspectives on designing socially acceptable social robots and robot avatars for Dubai and multicultural societies**

Dr. Laura Aymerich-Franch
Dubai Future Labs, Dubai Future Foundation, Dubai, United Arab Emirates
Corresponding author: laura.aymerich@dubaifuture.gov.ae
ORCID ID: 0000-0001-7627-5770

Dr. Tarek Taha
Dubai Future Labs, Dubai Future Foundation, Dubai, United Arab Emirates

Prof. Hiroshi Ishiguro
Department of Systems Innovation, Osaka University, Osaka, Japan

Prof. Takahiro Miyashita
ATR (Advanced Telecommunications Research Institute International), Interaction Technology Bank, Keihanna Science City, Kyoto, Japan

Prof. Paolo Dario
Dubai Future Labs, Dubai Future Foundation, Dubai, United Arab Emirates
The BioRobotics Institute, Scuola Superiore Sant'Anna, Pisa, Italy

**Competing interests**

The authors declare that they have no competing interests.

# Stakeholder perspectives on designing socially acceptable social robots and robot avatars for Dubai and multicultural societies

**Abstract**

Robot avatars for customer service are gaining traction in Japan. However, their acceptance in other societal contexts remains underexplored, complicating efforts to design robot avatars suitable for diverse cultural environments. To address this, we interviewed key stakeholders in Dubai's service sector to gain insights into their experiences deploying social robots for customer service, as well as their opinions on the most useful tasks and design features that could maximize customer acceptance of robot avatars in Dubai. Providing information and guiding individuals to specific locations were identified as the most valued functions. Regarding appearance, robotic-looking, highly anthropomorphic designs were the most preferred. Ultra-realistic androids and cartoonish-looking robots elicited mixed reactions, while hybrid androids, low-anthropomorphic robotic designs, and animal-looking robots were considered less suitable or discouraged. Additionally, a psycho-sociological analysis revealed that interactions with robot avatars are influenced by their symbolic meaning, context, and affordances. These findings offer pioneering insights into culturally adaptive robot avatar design, addressing a significant research gap and providing actionable guidelines for deploying socially acceptable robots and avatars in multicultural contexts worldwide.

## 1. Introduction

Robot avatars are hybrid interaction robots that can perform tasks autonomously, while also being controlled by a human operator [1–4]. Operators can interact socially through cybernetic avatars, control the avatar body, and communicate verbally through it [5–8]. Robot avatars for customer service have primarily been explored in Japan [3, 9–12]. For instance, this technology has been used in an avatar robot cafe for people of reduced mobility to be able to teleoperate a robot avatar waiter and interact with customers [13, 14]. Also, some experiences have looked at how robot avatars could potentially be used as salespersons [9]. However, deployments and acceptance of this emerging technology in other societal contexts remain largely unexplored, posing challenges to designing robot avatars that cater to diverse cultural environments.

Several studies have highlighted the importance of cultural factors in shaping human-robot interaction. Previous research suggests that cultural background significantly influences attitudes, perceptions, and acceptance of robots [15]. Cultural differences also affect robots' likeability, engagement, trust, and satisfaction [16]. Dubai, with its multicultural population and ambitious technological initiatives, provides a significantly different setting compared to Japan for the use of robot avatars. Notably, more than 200 nationalities live and work in the UAE [17]. Additionally, the need to create adaptive systems that can recognize and respond to cultural cues to enhance acceptance and interaction quality can become particularly challenging in highly multicultural societies like Dubai. Given the lack of prior studies specifically analyzing social robot and robot avatar acceptance for customer service in Dubai, the UAE, or other highly multicultural societies, this research addresses a significant gap in our understanding of human-robot interaction in this unique cultural context.

To address this gap, we conducted 10 in-depth interviews with key industry and government actors in the service sector in Dubai who had experience deploying social robots for customer service in their organizations. The goal was to gain insights into their experiences and explore which design features could help enhance robot avatar acceptance if this technology were introduced in Dubai for customer service. The organizations represented diverse sectors, including healthcare, transportation, retail, food and beverage, public services, robot developers, and business solutions providers with social robots.

The first part of the interview focused on gathering information about the organizations' experiences with social robot deployments. Social robots were introduced to participants as "robots designed to interact with humans using social cues". Specifically, we examined which robots were deployed, their tasks, customer receptiveness and perception, and the overall organizational evaluation of these deployments, to respond to the following question:

RQ1. What are the key experiences in Dubai's service sector with deploying social robots for customer service?

The design of social robots and robot avatars for the service sector demands a sophisticated approach that facilitates interactions but at the same time is able to mitigate unintended interruptions. For example, a social robot deployed in a shopping center should be designed to effectively guide customers while simultaneously discouraging interference from other visitors that might impede its functionality. Similarly, delivery robots need designs that not only appeal to customers but also discourage bystanders from obstructing their paths during delivery tasks. These scenarios highlight the importance of developing designs that can actively influence human behavior around robots. To design social robots and avatars that can effectively accomplish their tasks for assisting customers, it is critical to gain better insights on the factors that influence human behavior with social robots for customer service in real deployments. The following research question was formulated to examine this aspect:

RQ2. What psycho-sociological factors influence human interactions with social robots for customer service in real scenarios and how can these insights be leveraged to guide behavior?

To examine this question, we selected Symbolic Interactionism, Affordance Theory, and The Media Equation to guide our analysis. Symbolic interactionism is a sociological perspective that explores how individuals construct, interpret, and adjust social meanings through interaction and communication [18]. Central to this framework is the role of symbols, language, and shared meanings in shaping social behavior and the perception of reality. People enact roles within social interactions, often using objects, gestures, or appearances to convey meaning. For instance, items such as uniforms, luxury goods, or specific styles function as symbols of social status, authority, or group identity. These objects transcend practical use, serving primarily to communicate social meaning in various contexts [19]. Affordance Theory [20] states that living beings perceive the environment not only in terms of object shapes and spatial relationships but also in terms of object possibilities for action. In HCI, affordances refer to the perceived and actual properties of an object that determine how it can be used [21]. Finally, the Media Equation, from the field of Communication studies, suggests that people treat media and technology as if they were real social entities, applying human social norms and interactions to them [22]. These theoretical lenses provide valuable insights into the dynamic interplay between design elements and user behavior, particularly in contexts requiring robots to balance functionality with user engagement. This analysis revealed four critical themes that are essential for designing social robots and robot avatars to maximize acceptance. These themes, and their implications for design, are presented in the results section.

The second part of the interview introduced robot avatar technologies and various categories of robot avatar appearances. Robot avatars were introduced as "hybrid interaction robots that combine autonomous capabilities with teleoperated control. They can interact socially and perform tasks on their own, while also being controlled by a human operator for real-time interaction". Robot appearance is a key feature for acceptance that can also influence how humans interact and behave with robots [23]. We first inquired about the anticipated customer acceptance of different robot avatar appearance categories. Subsequently, we invited interviewees to brainstorm the ideal robot avatar design for their organization. In particular, we sought to provide an answer to the following research questions:

RQ3. What are the most meaningful tasks for robot avatars in Dubai's service sector, what challenges could they help resolve in customer service, and where should they be placed?

RQ4. Which design features, particularly related to appearance, could potentially enhance the acceptance of robot avatars in Dubai's multicultural environment?

## 3. Procedure

We employed a purposive sampling strategy to select interview participants. Initially, organizations within Dubai's service sector with documented experience in deploying social robots in real-world scenarios were identified through press releases and professional network. Subsequently, we contacted the organizations to identify the person who was most directly involved in overseeing the deployment/s and asked them to participate in the study. In total, we conducted 10 semi-structured in-depth interviews with key industry and governmental stakeholders in the service sector in Dubai, all of whom had experience deploying social robots for customer service in their respective organizations. Two of the interviewees had additionally been involved in deploying social robots at Expo 2020 Dubai, and their experiences from this event were also gathered. Participant organizations were drawn from various sectors, including healthcare, transportation, food & beverage, public services, retail and hospitality, robot developers, and business solutions providers with social robots. Each interview lasted approximately 50 minutes. Interviews were conducted either online or in person, and all interviews were recorded locally using an internal voice recorder. The recordings were transcribed manually for analysis. Prior to the interview, participants provided written informed consent. Participants' profiles and industries (Table 1A) as well as the full list of interview questions can be found in the supplementary material.

## 4. Results

### 4.1 Social robot deployments in Dubai

In response to RQ1, we asked the interviewees to share their experience with deploying social robots for customer service sector within their organizations. All participating organizations had experience deploying social robots for customer service, although the duration of these experiences varied significantly, ranging from 2-day deployments to over seven years and still ongoing. The social robots deployed included Pepper (Softbank), Promobot (Promobot), Ava (Ava Robotics), Temi (Robotemi), Sota (VStone), food delivery robots from Pudu, Nao (Softbank), various robots from Terminus Group (deployed at Expo 2020), robot waiters (manufacturer undisclosed), a check-in robot (internal R&D with external partners), and delivery and food-serving robots (internal R&D with external partners).

The primary tasks performed by the social robots included providing information, offering guidance, delivering entertainment, and managing check-in processes. In hospitality contexts, food delivery was the robots' primary function.

Expo 2020, held in Dubai, marked one of the largest deployments of social robots, with over a hundred robots in operation. As one of the interviewees with experience deploying in this event explained, "This was the first time we were bringing the robots to the public to interact with them, as opposed to having them as part of an exhibition. We deployed them for six months." At Expo 2020, robots performed various tasks, including guiding visitors, patrolling for security, providing maps of the Expo site, and entertaining visitors, particularly children, through a robot mascot.

Shorter deployments shared by the interviewees were typically part of pilot tests or targeted campaigns. For example, one organization deployed a social robot "to distribute gifts at a mall during a National Day campaign. People loved it, especially the kids."

### 4.2. Psycho-sociological factors influencing behavior with robots

RQ2 examined what psycho-sociological factors influence human interactions with social robots for customer service in real scenarios and how these insights can be leveraged to guide behavior.

An analysis of stakeholders' experiences with social robot deployments, framed through the lenses of Symbolic Interactionism, Affordance Theory, and The Media Equation, revealed a series of key findings worth considering for an effective design of social robots and robot avatars.

*Theme 1. Robots are a unique category in the collective imaginary distinct from machines, which comes with certain design expectations*

In line with Symbolic Interactionism, our findings suggest that robots occupy a unique category in the collective imaginary, distinct from other machines, which is associated with certain design expectations with regard to appearance. In particular, they are perceived as social entities which, despite being mechanical, are expected to possess at least minimal anthropomorphic features, enabling them to acknowledge and communicate with humans through these human-like traits: "The delivery robots have eyes, this is because people want it. They expect an amicable characteristic. Once you add a couple of eyes it is not a machine anymore, you achieve this

purpose. Adding this feature is also conveying the message that 'I am not a car, I won't run over you'."

This distinction might also help explain why certain appearances are perceived as more socially acceptable than others by the participants. In particular, the category of *robotic looking, highly anthropomorphic* was the most closely aligned with the mental representation of a robot in the collective imaginary of the Dubaian society. Some interviewees identified this category as the most aligned with social expectations: "This is the one that gives me the best impression. The type that I would expect." These designs enable robots to adopt roles and meanings that resonate with socially shared representations of the concept 'robot', reinforcing their place as social entities capable of interaction and communication.

From the lens of the Media Equation perspective, our findings further suggest that robots epitomize the concept that people treat media and interactive technologies as social entities. Social robots represent a particularly relevant form of this phenomenon. Their design expectations are heightened precisely because people not only perceive them as tools but as social actors. Thus, robots are at the forefront of interactive technologies that humans want to engage with socially, and as a result, the demand for social elements in their design is particularly pronounced.

*Theme 2. Context shapes behavior*

In line with Gibson's theory of affordance [20], our findings suggest that people's behavior and interactions with robots is context-dependent. In other words, the environment influences how individuals perceive and engage with the robot. The same robot can elicit varying behaviors depending on the affordances provided by its surrounding environment.

One of the organizations we interviewed had experience deploying the same delivery robot in two distinct contexts: Expo 2020 and a Dubai residential area. They observed how people's behaviors with the same robot changed significantly depending on the context:

> Interviewee: People wanted to interact a lot with the delivery robot during Expo. Play, speak to it – they assumed the robot would reply just because it is a robot. People have an expectation for the robot to speak to them. They tried to interact with it, touching the screen, trying to give it orders. People of all ages, not only kids.
>
> Researcher: Did it happen often that people prevented the delivery robot from achieving its delivery task?

> Interviewee: At Expo, yes, it was always happening. For this reason, we had a voice recording. The robot was verbally announcing that it was on a way to deliver so that people moved out of the way but people stopped them. The strategy that worked best was to stop the robot until the person got bored of it. If the robot tried to sidestep the person then the person kept playing with it.

In contrast, when the robot was deployed in one of the residential communities in Dubai:

> Interviewee: People behaved differently. They played less with it. They understood it is a service for delivery. People liked it but they attempted less to interact with it. That was a real experience of delivery that lasted a year and a half, not an experiment.

*Theme 3. Interactions with robots are shaped by their affordances*

In line with the approaches to affordance theory in HRI studies [21], our findings suggest that interactions with robots are heavily influenced by their affordances—the perceived action possibilities presented by the robot in relation to the user. Affordances shape how individuals understand and engage with robots.

Robots with unclear affordances were viewed as less suitable for social contexts, reinforcing the necessity for designs that somehow communicate their purpose and functionality. This finding in particular helps explain why robots with low-anthropomorphic features received very poor evaluations overall. As one of the participants noted: "It is not clear how to interact, it's not user-friendly". Another participant mentioned about the designs representing the low-anthropomorphic category: "this looks like a telephone, is not easy to interact with it, for social interaction is not good". It also contributes to explain why robotic designs with anthropomorphic traits were perceived as more suitable, as they more clearly communicated how to engage.

*Theme 4. Interactions with robots are shaped by symbolic meaning*

In line with Symbolic Interactionism, our findings indicate that human-robot interactions are shaped not just by the robot's functionality, but also by the symbolic meanings it conveys through its appearance and behavior. Just as people assign roles and significance to objects and symbols in social interactions, they do the same with robots.

Participants highlighted how symbolic cues affected user behavior. As one of the interviewees explained: "One of the robots had minimal functionality but its appearance was very friendly and cartoonish. Kids immediately started to run towards the robot and climb it." Conversely, a patrolling robot with a more authoritative appearance evoked hesitancy: "We had a patrolling robot that was more like a police type and people were scared to go closer".

**4.3. Challenges, tasks, and spaces for robot avatars**

RQ3 investigated the most meaningful tasks for robot avatars in Dubai's service sector, as well as the challenges that robot avatars could help resolve in customer service, and the suitable spaces to place these robotic entities. In response to this question, interviewees identified specific customer needs and challenges that could be addressed using robot avatars.

One of the challenges was reducing staff workload during peak periods, special campaigns, or events, which was common across several sectors. Robot avatars were proposed as a solution, offering services such as wayfinding, providing information, personalized recommendations, delivering objects or food within premises, and entertaining customers during waiting periods.

Multilingual functionality was highlighted as another key feature by several interviewees, due to Dubai's multicultural nature and its significant international tourist traffic. An interviewee also noted the importance of supporting customers who might struggle with existing services: "Regular customers know how to use the services, but some may face difficulties. For example, language barriers, individuals with autism, or people of determination may require additional support. Intelligent robots could assist by providing multilingual capabilities, carrying luggage, and more. This would reduce staff requirements and improve productivity."

Additionally, one participant mentioned the potential for robots to serve as smartphone charging stations.

Finally in the library robot avatars were found a useful technology to help locate books, provide recommendations and summaries, and assist with research activities.

Interviewees were additionally asked to think about their customer needs and identify or rate the ideal tasks for a robot avatar in their organization considering a pre-defined list of tasks (see Table 1). Providing information and offering directions to help individuals locate specific places were the most valued functions: "We have over 90.000 products in our stores, many of them our customers don't know about. If the robot could explain about them it would benefit us a great deal". Customer registration and check-in, tele-presence, patrolling, object delivery, gathering survey data, collecting customer feedback and handling complaints were considered particularly valuable by half or more of the participants. While these functions were considered relevant for customer service in these organizations, lower ratings by other interviewees were often due to the perception that the robot was unsuitable for the task or that a more effective system was already in place. For example, in the case of security, one participant noted, "because we already have many cameras" or "it is not much of a necessity because this is a very safe country", or in the case of transportation and delivery: "the robot would be too slow, it is faster to use a human". Gathering feedback and handling complaints in particular were rated either very high or very low. While all interviewees agreed that this was an important task, some were particularly reluctant to delegate this function to a robot as they felt that this responsibility was too significant to be delegated to a machine. One participant remarked, "I don't think the robot would be good at managing people's emotions. They would prefer a human for that," while another expressed concern that "they might feel that they are not being taken seriously." Goods or food transportation and entertainment was only relevant to smaller number of interviewees. However, for one of the organizations, food and object transportation was the most and only relevant function due to this being the core of their business. Two participants suggested adding multi-language functionality to facilitate interaction with a diverse audience, which could be particularly meaningful in spaces such as the airport.

| Participant | Inform. | Guidance | Complain. | Registration | Data collec. | Patrol. | Telepr. | Delivery | Transport | Entert. |
|---|---|---|---|---|---|---|---|---|---|---|
| P1 | Yes | Yes | No | Yes |  | No | Yes | No | No | No |
| P2 | Yes | Yes | Yes |  | Yes | Yes |  | No | No |  |
| P3 | 7 | 6 | 1 | 4 | 5 | 7 | 1 | 5 | 5 | 8 |
| P4 | Yes | Yes | Yes | Yes | Yes | Yes | Yes | Yes | Yes | Yes |
| P5 | 10 | 10 | 3 | 8 | 2 | 5 | 6 | 5 | 5 | 5 |
| P6 | 10 | 10 | 10 | 8 | 10 | 7 | 10 | 10 | 5 | 6 |
| P7 | 10 | 10 | 8-9 | 2-3 | 8-9 | 2-3 | 10 | 2-3 | 2-3 | 2-3 |
| P8 |  |  |  |  |  |  |  | Yes | Yes |  |
| P9 | 10 | 7 | 8 | 8 | 8 | 7 | 6 | 7 | 7 | 5 |
| P10 | 10 | 10 | 10 | 5 | 5 | 1 | 10 | 10 | 10 | 8 |
| Relevance Count* | 9/10 | 8/10 | 6/10 | 5/10 | 5/10 | 5/10 | 5/10 | 5/10 | 4/10 | 3/10 |

Table 1. Key tasks for social robot avatars in customer service settings
*Specifically mentioned as relevant or rated 7 or higher*

Finally, regarding spaces, main lobby in hospitals, metro stations, restaurants, shopping malls, cinemas, airports, libraries, and staff offices were mentioned among the potential spaces suitable for robot avatars.

### 4.4. Acceptance of social robot avatars

RQ4 examined which design features, particularly related to appearance, could potentially enhance the acceptance of robot avatars in Dubai's multicultural environment.

Overall, interviewees expressed varied but generally optimistic views toward the acceptance of social robot avatars among their customers: "They would love it!".

A participant who had experience deploying multiple social robots highlighted two key aspects that could determine acceptance: height - less tall than a human- and appearance - humanoid but not android-. In particular, this organization found Pepper the most ideal robot among all available options: "When it comes to acceptance, why did we find Pepper to be a better robot compared to other robots out there? It's because it wasn't an intimidating robot. It is designed to be 4 feet, it looks up to you, more child-like so it does not scare you, it has a sweet voice that is not scary. (…) Some robots are 5 or 6 feet tall. Imagine one of these robots coming after you, it would be scary. And if it looks too human (…) it's a bit on the creepy side. Pepper wasn't too tall. Pepper looks up at you, it gives you respect".

Participants emphasized the importance of designing social robots and avatars tailored to Dubai's unique needs, particularly its multicultural environment. They suggested features that cater to diverse languages and cultural backgrounds: "We are becoming a lot more diverse as a community in the UAE. We need to cater to more languages, cater to more cultures, to tourists. This is a way how a technology like this could go well." Furthermore, understanding Dubai's societal context, distinct from regions where robots are widely deployed, is critical for addressing real needs: "The difference between the rest of the world where these robots do well and here is that in these other places they do well because there is a shortage of labor. I have seen this in Japan, Singapore, Hong Kong… where you have a shortage of labor or the cost of labor is much higher. Here we don't have this limitation; we have enough people working in hospitality."

To find a niche for robot avatars, it is also important to understand and acknowledge what tasks a robot can do well and what tasks a human can do better, and combine their capabilities. As a participant pointed out: "There are a lot of repetitive tasks, explaining features, explaining what a product can do, etc. that a robot can do well, a robot doesn't get tired, so there are certain things that it can do well, and humans do certain things well. It is a matter of combining abilities to see where the robot can do the job and then it will pass it on to the human"

Additionally, several participants highlighted the fact that the robots' ability to successfully perform relevant tasks was critical for acceptance, with functionality consistently emphasized as more important than appearance. One interviewee noted, "Acceptance increases depending on the functions it can successfully perform. Probably more than an 8, but it is not just about appearance, we need to make sure the information is provided successfully." Another remarked, "I have no reason to think it would not be accepted. But functionality is critical, more than appearance; if it is not fulfilling its function properly, it won't be accepted." Initial enthusiasm for robots was also highlighted, but several interviewees warned that this excitement might diminish if the robot failed to provide essential information or services effectively. As one explained, "People are initially interested in the robot, but the initial excitement lowers if the robot is not able to give you the information you need as a customer." While the appearance of the robots was a key point of discussion, these insights underline that functionality and task relevance are crucial for sustaining long-term acceptance of social robot avatars. As one participant emphasized, "The value is in the information or service the robot provides. If the robot can interact and provide valuable information, it would be well-received. The value for the patient is on the service, not on the look." This perspective underscores the importance of ensuring that social robots are not only visually appealing but also effective in fulfilling their intended tasks.

### *4.4.1. Social robot avatar appearances*

To explore customer preferences regarding the appearance of social robot avatars, interviewees were shown various robot avatars with different levels of human-like features and were asked to discuss and/or rate the expected level of customer acceptance for each type. The six types of robot appearances presented included (Fig. 1): ultra-realistic androids, which closely resemble human beings with skin-like textures; hybrid androids, which feature a mix of human and robotic characteristics; robotic-looking, highly anthropomorphic robots, which have body shapes and faces reminiscent of humans but clearly robotic in nature; robotic-looking, low anthropomorphic robots, which have minimal human-like features; cartoonish-looking robots, characterized by exaggerated, playful features; and animal-looking robots, designed to resemble animals. Participants were asked to consider each robot's potential acceptance by their customer base based on appearance and indicate their preferences, either rating from 1 to 10 or selecting the ones they found more suitable.

The category of robotic-looking, highly anthropomorphic robots—featuring humanoid shapes with clearly robotic elements—emerged as the most accepted and preferred by most interviewees. Respondents noted that these robots strike the right balance between being friendly and non-threatening, without crossing into the discomfort zone of ultra-realistic androids: "It is playful, human-like, not too mechanical, but also not creepy, it looks safe, and it's less creepy".

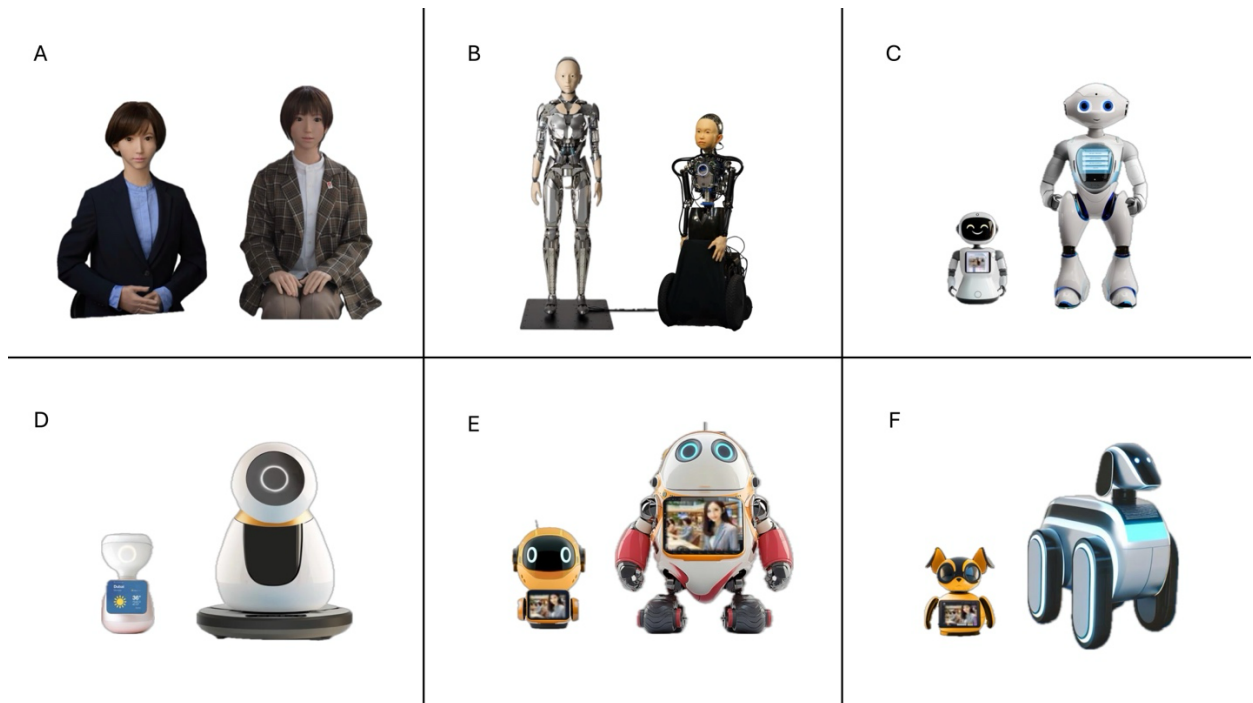

*Fig. 1 Robot avatar appearances: Ultra-realistic androids (A), hybrid androids (B), robotic-looking, highly anthropomorphic robots (C), robotic-looking, low anthropomorphic robots (D), cartoonish-looking robots (E), animal-looking robots (F)*

A key factor in their acceptance was familiarity with this type of robot, which participants often described as what was expected: "This is the one that gives me the best impression—the type I would expect." and familiar: "People are familiar with this (…). It would be easy to interact with, and it's friendly and approachable". Interviewees also noted that this design was well-established, and already proven to work: "I prefer this one the most. It's familiar to what we already have, visitors already like it, and it's easy to interact with."

Additionally, interviewees frequently emphasized that this type of robot would appeal to a broad range of users, from children to adults: "Looking at the demographics we serve, from kids to adults, it seems the most neutral one." Overall, this type of design was described as safe, easy to interact with, and suitable for a wide range of functions and audiences, from providing information to offering entertainment.

In contrast, robotic-looking, moderately anthropomorphic robots, which have minimal human-like features, were generally rated low. This category was seen as too simplistic, lacking capacity to engage customers. Overall, interviewees felt that these robots would not be effective in customer service or social interaction scenarios due to their overly basic design and inability to connect with users on an emotional level. These robots were considered more suitable for utilitarian purposes, such as B2B applications, rather than customer-facing roles that require interaction and engagement: "If it's like in your office, for B2B ok, but for customer service doesn't seem the best, it looks too basic".

Ultra-realistic androids raised particularly divergent opinions. On one hand, half of the respondents viewed them favorably, and it was selected as the preferred appearance by three organizations. The interviewees appreciated the potential for customization to align with cultural preferences, noting that such robots could be designed to better fit specific demographic contexts. For example, one interviewee suggested that ultra-realistic androids could be adapted to Middle Eastern cultural norms, such as wearing traditional attire like an abaya or kandura. However, despite these positive comments, the other half of interviewees expressed particular discomfort with this robot type, attributable to the "uncanny valley" effect, in which the robot's appearance is almost human but still not quite right, creating an unsettling experience. In particular, the ambiguity of the robot's appearance, where it was unclear whether it was human or not, also led to negative reactions. As one participant remarked, "They are not perfectly human-looking, but they are not robots either." Another noted, "I don't see this one working well in Dubai. In Dubai, things related to the future have a futuristic appearance. This robot is confusing. Is it a robot or not? There has to be a clear difference between robot and human and it needs to be clearly identifiable as such." Some respondents also noted that in certain cultural contexts, such as in Arab cultures, ultra-realistic robots might be perceived as eerie or even creepy.

In contrast, hybrid androids, which blend human and robotic features, were consistently rejected. Many participants found hybrid androids unsettling, with visible mechanical components such as exposed wires contributing to the impression that these robots were unfinished or unrefined. As one interviewee noted, "I would never create a robot with cables outside, it seems unfinished, as if you were doing an experiment. You need to create the impression that it is a service, fully-functioning, and fully-finished." Respondents cited that this hybrid appearance "does not look friendly for the visitors, a bit scary" Ratings for hybrid androids were consistently low, with many interviewees expressing that such robots would struggle to facilitate natural human-robot interaction, especially in customer-facing roles.

The cartoonish-looking robots category received mixed responses. This type of appearance was particularly well-received by half of the sample. For one of the interviewees, cartoonish was the ideal appearance: "considering that our previous experience with deployments belongs to this category, and how well it was received, something cartoonish would be very well accepted". Overall, respondents appreciated the clear and unmistakable "robotic" look of these designs, which would make it easy for users, especially children, to understand the robot's role. However, the other half of the sample expressed reservations about their suitability for adult users, noting that such designs might be perceived as childish or unsuitable in certain settings, such as way-finding. While these robots were seen as an excellent option for environments aimed at younger audiences, their appeal for adult customers was questioned: "Kids would like it but it's not for everybody, this would work to interact with kids, but adults want to see something else".

Finally, animal-looking robots was overall rated very poorly. While, participants acknowledged their potential in child-focused environments, where their playful and familiar appearance could be engaging, they did not find them suitable for adults and customer service roles: "this is more targeted for kids, it would look more like a toy than a helper". Additionally, a respondent expressed concern that using an animal-like robot for delivery could evoke negative feelings, especially for pet owners: "It is like slavery for animals, I don't like it, if I had a dog I would get angry to see that a dog is used to do delivery". Table 2 shows a summary of the results.

| Participant | Android | Robotic, Highly Anthrop. | Robotic, Low Anthrop. | Hybrid Android | Cartoonish-looking | Animal-looking |
|---|---|---|---|---|---|---|
| P1 | **Preferred** | Second preferred | - | - | - | - |
| P2 | - | **Preferred** | - | - | - | - |
| P3 | 3-4 | **8-9** | 6 | 2 | 7 | 6 |
| P4 | 3 | **7** | 5 | 1 | 6 | 2 |
| P5 | **8** | **8** | 1-2 | 5 | 7-8 | 7-8 |
| P6 | 8 | **9-10** | 3 | 1 | 8 | 3 |
| P7 | **7-8** | 6-7 | 2-3 | 3-4 | 3 | 3 |
| P8 | 3 | **10** | 8 | 5 | 8 | 7 |
| P9 | 6 | **9** | 7 | 4 | 6 | 5 |
| P10 | 10 | 10 | 8 | 4 | **Above 10** | 10 |
| Acceptance count* | 5/10 | 9/10 | 3/10 | 0/10 | 5/10 | 3/10 |

Table 2. Preferred robot avatar appearances

*Acceptance is defined when rated 7 or higher or preferred option. Preferred option by participant is indicated in bold.*

### 4.4.2. Other design considerations

Regarding materials, most interviewees agreed that plastic was the best material, some added silicon might also be suitable. For those that whose preferred option was an android, skin-like texture was the chosen material. Table 3 illustrates the results by participant.

| Participant | Skin-like Texture | Plastic | Silicone | Metal | Comments |
|---|---|---|---|---|---|
| P1 |  | Preferred (better for maintenance and cleanliness) |  | No | Sanitizable materials is a must |
| P2 |  | Preferred |  |  | Materials cannot get dirty |
| P3 |  | Preferred |  |  |  |
| P4 |  | Preferred |  |  |  |
| P5 | Preferred |  |  |  |  |
| P6 |  | Preferred | Preferred |  |  |
| P7 | Preferred |  |  |  |  |
| P8 |  | Preferred | Preferred (when there is direct interaction with the person) |  | Materials cannot hurt the person |
| P9 |  | Preferred | Preferred |  |  |
| P10 |  | Preferred (to not scare people) |  |  |  |
| Preference count | 2/10 | 8/10 | 3/10 | 0/10 |  |

Table 3. Preferred materials for robot avatar designs

Regarding size, participants 'preferences ranged from 40 cm to 1.65 m. Most favored sizes between 1.2 m and 1.5 m when considering robotic designs, whereas preferences shifted to 1.5 m to 1.6 m when envisioning android designs, in line with the human average size. Table 4 contains the results by participant.

| Participant | Height preferences |
|---|---|
| P1 | 1.5-1.6m |
| P2 | 1.5m |
| P3 | 1-1.2m |
| P4 | 1.2m |
| P5 | 1.2-1.5m |
| P6 | 1.5m (for interaction with adults, but it depends on location: children or adults areas, should be catered to that) |
| P7 | 1.5 – 1.65m (like a person) |
| P8 | 1.2 – 1.5m |
| P9 | 1.2m |
| P10 | 40cm |

Table 4. Height preferences for robot avatar designs

Regarding degrees of freedom, most participants expressed a preference for high mobility in the robot avatar design. Participants who favored android designs, in particular, preferred the robot to closely imitate the full range of human movements. One participant remarked, "even the eyes," emphasizing the importance of detailed mimicry. However, some other participants prioritized fewer degrees of freedom to not jeopardize functionality: "we prioritize speed and functionality. More DOF might make those more challenging." Also, another interviewee noted that less DOF were preferred "so it does not distract customers from getting the information with excessive movement": Table 5 contains the results by participant.

| Participant | Mobility – DOF preferences |
|---|---|
| P1 | Both arms and legs - so that it looks more human-like |
| P2 | As many parts as possible - as human-like as possible |
| P3 | Low degrees of freedom preferred |
| P4 | Head and body or head, body, and arms |
| P5 | As many DOF as possible |
| P6 | Head, body, and arms |
| P7 | Realistic movement - like humans do |
| P8 | At least hands and head |
| P9 | Head and body (for easier deployment) or head, body, and arms (more engaging but causing more issues for reparations) |
| P10 | Head only (to not be distracting) |

Table 5. DOF preferences for robot avatar designs

When asked about preference for the robot avatar shape, most interviewees preferred rounded shapes, emphasizing their approachable and non-threatening appearance. One participant remarked that "rounded shapes give a better impression and appear less threatening." A minority of participants preferred squared designs, particularly in non-android contexts, while one expressed no strong preference for either shape. Table 6 summarizes the results by participant.

| Participant | Preferred shape design |
|---|---|

| P1 | Rounded |
|---|---|
| P2 | Squared |
| P3 | No preference |
| P4 | Rounded |
| P5 | Rounded |
| P6 | Rounded |
| P7 | Squared |
| P8 | Rounded or squared but with rounded edges |
| P9 | Rounded |
| P10 | Rounded |

Table 6. Shape preferences for robot avatar designs

A majority of interviewees preferred limbs designed with hands and fingers, some citing their usefulness for more tasks: "Hands and fingers are preferred because they can perform more functions." However, a minority of participants mentioned it would depend on the function, whichever could work best based on the function. Table 7 summarizes the results by participant.

| Participant | Preferred upper limb design |
|---|---|
| P1 | Hands with fingers |
| P2 | Hands with fingers |
| P3 | Functional (whichever works better on each case) |
| P4 | Hands with fingers |
| P5 | Hands with fingers |
| P6 | Hands with fingers |
| P7 | Robotic arm without fingers |
| P8 | Functional (whichever works better on each case) |
| P9 | Hands with fingers |
| P10 | Hands with fingers (as long as it is not scary) |

Table 7. Limb design preferences for robot avatars

Most participants preferred mobile robots, whether because "it is more appealing to customers" or it allows the robot to do "more functions, such as transporting objects". A participant suggested that ideally it would have the two modalities: "A robot that had both options and you could adapt depending on what's happening around, for instance, if it's crowded you can keep it static." Another participant stressed that mobility opportunity depends heavily on the robot's function, explaining, "It depends on the function. For security and patrol, it might be a good use case to use a mobile robot. But when we tried delivery with a robot, the speed was so slow that it was not feasible. It depends on the use case we look at." Table 8 summarizes the results by participant.

| Participant | Mobility Preferences |
|---|---|
| P1 | Mobile |
| P2 | Mobile |
| P3 | Depends on function |
| P4 | Mobile |
| P5 | Mobile |
| P6 | Mobile |
| P7 | Mobile and Static (adaptable depending on circumstances) |
| P8 | Mobile |

| Participant | |
|---|---|
| P9 | Mobile |
| P10 | Stationary |

Table 8. Mobility preferences for robot avatar designs

A majority of participants preferred physical robots over virtual avatars or on-screen representations, citing their added value compared to the ubiquitous presence of screens. One participant emphasized, "Physical because a tablet everyone has one, if it has movement and can talk you add another dimension to the experience." Another noted, "A static one and a screen is more or less the same…mobile better," reinforcing the sentiment that physical robots bring unique interactivity that screens lack. Additionally, participants highlighted the engaging and futuristic appeal of physical robots, with one stating, "A robot in a screen is not representing the future. It is not the way you really want to interact. It loses power." However, one of the participants acknowledged that on-screen avatars might be suitable depending on the use case.

| Participant | Preferred modality: physical vs virtual |
|---|---|
| P1 | Physical |
| P2 | Physical |
| P3 | Depends on the use case |
| P4 | Physical |
| P5 | Physical |
| P6 | Physical |
| P7 | Physical |
| P8 | Physical |
| P9 | Multiplatform |
| P10 | Physical |

Table 9. Modality preferences for robot avatars

A majority of participants preferred verbal interactions for autonomous robots. One participant explained, "Voice because the action is faster, typing takes time, and some people might not have the ability to interact with displays. If voice did not work, you could have the display as a back-up plan." Another participant highlighted the significance of verbal interaction in a multicultural context, citing the "added value of multi-language for interacting with patients of different nationalities." However, one participant prioritized the use of a screen over voice interaction, stating, "A mix of both would be more ideal, but if I could only choose one, it would be touch-screen to accommodate a more diverse audience and otherwise maybe sometimes the robot would not understand." Table 10 summarizes the interaction preferences during autonomous mode by participant.

| Participant | Preferred Interaction Mode |
| --- | --- |
| P1 | Verbal |
| P2 | Both combined |
| P3 | Both (verbal more critical, screen to support) |
| P4 | Both combined |
| P5 | Verbal |
| P6 | Both (but screen more critical) |
| P7 | Verbal |
| P8 | Both, but verbal more critical |
| P9 | Both combined |
| P10 | Both combined |

Table 10. Interaction preferences during autonomous functioning

Participants were asked whether, in tele-operated interactions, they would prefer the operator to speak through the robot or appear on a screen. Most participants preferred that tele-operators speak through the robot, with the opinion that their customers might not like seeing the tele-operator on the screen as it might defeat the purpose of having a robot: "The aim of the robot is to avoid having people in front of you, so a video call defeats the purpose." However, two participants preferred the use of a screen. One participant, particularly considering android designs, emphasized the importance of distinguishing between autonomous and tele-operated interactions: "otherwise, it seems possessed.". Finally, one of the participants pointed at the fact that verbal interactions with the robot in crowded environments are sometimes challenging as the robot is not able to properly capture the audio which causes difficulties in communication and suggested it might be best in these scenarios to offer a different solution, such as interacting with the teleoperator on the phone instead. Table 11 summarizes the results by participant.

| Participant | Preferred Interaction Mode |
|---|---|
| P1 | On the screen |
| P2 | Through the robot |
| P3 | Through the robot |
| P4 | Through the robot |
| P5 | Through the robot |
| P6 | Through the robot |
| P7 | Through the robot |
| P8 | On the screen |
| P9 | Not with robot, on the phone |
| P10 | Both options are good |

Table 11. Interaction preferences during tele-operated mode

Participants were asked how they preferred robots to express emotions, considering the options of facial expressions, body movement, sounds, and emoticons. The majority emphasized the use of body movement and facial expressions as the preferred option, or all the options combined. A participant suggested specific constraints for sounds: "Probably no sound because we have lots of announcements in the stations, so it would not work well." Only one of the participants had a preference for the robot not showing emotions, particularly facial expressions. Table 12 shows the results by participant.

| Participant | Preferred Emotional Expression |
|---|---|
| P1 | Body movement and facial expressions |
| P2 | Body movement and facial expressions, no sound |
| P3 | All |
| P4 | All |
| P5 | All |
| P6 | No emotion expression, particularly no facial expressions |
| P7 | All |
| P8 | All (movement depending on the case) |
| P9 | All |
| P10 | Facial expressions |

Table 12. Emotion expression preferences for robot avatar designs

Participants were also asked about their preferred color combinations for robot avatars, with options including white, black, silver metal, pink copper metal, pale golden metal, and sustainability colors. White was often preferred. This color was perceived less tiring: "Discrete colors are better, or it can be tiring. That is why most of them are white. I would not do black or any tiring color." It was also associated with cleanliness: "Mainly white, because it is linked with sanitization and clean". Table 13 summarizes the results.

| Participant | Preferred Color Combination |
|---|---|
| P1 | Mainly white with branding colors and logo |
| P2 | White and silver metal |
| P3 | White |
| P4 | White, black, silver metal, pale golden metal |
| P5 | Customizable depending on context |
| P6 | Sustainability colors (to represent the values we stand by) |
| P7 | Skin-neutral (for android), silver metal (for others) |
| P8 | Discrete colors (white preferred) |
| P9 | White |
| P10 | White and silver metal |

Table 13. Color combination preferences for robot avatar designs

## 5. Discussion

This study aimed to explore stakeholder perspectives on the acceptance of social robots and robot avatars within Dubai's multicultural service sector. The objective was to understand prior experiences with social robot deployments, uncover psycho-sociological factors that might influence how people behave with robots when they are deployed for customer service in real environments, identify challenges and tasks within the service sector that would benefit from robot avatars, and evaluate design features to maximize acceptance from customers.

The study involved 10 in-depth interviews with stakeholders from diverse sectors, including healthcare, transportation, retail, food and beverage, public services, robot developers, and business solutions providers with social robots. These participants were selected based on their direct involvement in deploying social robots for customer service in Dubai.

In relation to RQ1, we found that social robot deployment experiences in Dubai are primarily used for tasks such as providing information, guiding individuals, facilitating check-ins, and offering entertainment. Notable examples include large-scale deployments during Expo 2020, as well as ongoing experiences spanning over seven years.

A psycho-sociological analysis of customer behavior with robots based on the insights provided by the participants from real deployment experiences in Dubai revealed four key considerations (RQ2). First, robots occupy a unique category in the collective imaginary, distinct from machines, and are expected to have at least minimal anthropomorphic features that facilitate human interaction. Second, behavior and interactions with robots are context-dependent, with the same robot eliciting different responses based on the environment. Third, robot affordances—how they convey action possibilities—strongly influence user interactions. Anthropomorphic designs were preferred for their clear communication of engagement methods, while low-anthropomorphic designs were often deemed unsuitable for social contexts. Forth, interactions are shaped by the symbolic meaning conveyed by robots' appearances and behaviors.

Individuals make decisions influenced by the symbolic meanings attached to objects, not solely their utility. Objects often carry meanings beyond their practical application, serving as markers of identity, status, or values. For example, luxury goods are frequently purchased to signal wealth or social standing rather than for their functional purpose. Similarly, organizations and individuals may choose to deploy robots in service environments not only for their operational efficiency but also for the symbolic meanings they convey (e.g., a reflection of innovation), adding intangible value that resonates with customers and stakeholders. Therefore, it is crucial to consider the symbolic implications of robot designs and evaluate whether the associated meanings align with the goals and values of the target audience or organization. By addressing these factors, designers can create robot avatars that are not only functional and engaging but also culturally and symbolically relevant, enhancing acceptance and integration within diverse service settings.

Overall, our findings underscore the importance of promoting a multidisciplinary approach by integrating sociological, psychological, and Communication theories into design processes, offering solutions grounded in a deeper understanding of human behavior.

In relation to robot avatar tasks (RQ3), interviewees indicated that if social robot avatars were deployed in their organizations, providing information and guidance would be the two most desired functions.

Finally, regarding design considerations (RQ4), robotic-looking, highly anthropomorphic designs were most preferred due to their balance between familiarity and a non-threatening appearance. Ultra-realistic androids elicited mixed reactions. Half of the sample selected them as their preferred appearance while the other half expressed notable rejection and discomfort. Several participants also stressed the importance of functionality over appearance, emphasizing that a robot's ability to perform tasks effectively was key to sustained acceptance. Plastic was widely preferred as a material for its hygienic properties and ease of maintenance. Ideal robot heights ranged from 1.2 to 1.5 meters for robotic designs, with slightly taller preferences for android designs to match the average human size. Rounded shapes were generally favored for their approachable and non-threatening qualities. Hands with fingers were preferred over no fingers for their functional versatility. Most participants emphasized the importance of mobility, viewing mobile robots as more versatile and engaging for users. In line with the previous, physical robots were overwhelmingly favored over virtual avatars, as virtual avatars where not seen as providing an added value, given the ubiquitous presence of screens in the society. Verbal interaction during autonomous mode was preferred, but ideally supported by screen interactions. For tele-operated interactions, participants preferred that operators speak through the robot rather than appear on a screen. In terms of emotional expression, body movement and facial expressions were the most preferred modes. White was the most preferred color for robot designs, linked to its association with cleanliness and a non-tiring appearance.

In sum, interviewees conveyed predominantly positive perspectives on the acceptance of social robot avatars for customer service within the multicultural context of Dubai. Nevertheless, several participants underscored that the robots' ability to effectively perform relevant tasks was essential for gaining acceptance, in addition to appearance.

By gathering insights from key industry and governmental actors with experience in deploying social robots, this research provides pioneering perspectives on robot acceptance, design preferences, and potential applications in Dubai's service sector. These findings are not only crucial for informing future robot deployments and design strategies in the region but also offer insights into how robot avatars can be designed and deployed to meet the demands of complex multicultural environments. Additionally, this study is part of the Japanese Moonshot Program [24], which focuses on pioneering innovative technologies to shape future societies. As part of this collaborative initiative, we are actively deploying social robots and avatars for customer service in Dubai, leveraging real-world environments to explore and refine social robot interactions. The insights gathered from stakeholders in this study are directly relevant to the ongoing development of the robot concept within the Moonshot framework. By integrating stakeholder feedback into the design process, we aim to ensure that the resulting social robots and robot avatars align with the cultural, social, and functional needs of Dubai's multicultural environment, ultimately enhancing customer acceptance and service efficiency.

Finally, by emphasizing a user-centered design that includes all relevant stakeholders from the initial stages of the process, the study provides a framework for optimizing the integration of robot avatars in the service sector and the society in general. The findings are intended to inform the design and deployment of robot avatars in similar multicultural contexts globally.

## 6. Data Availability Statement

The authors confirm that all data generated or analyzed during this study are included in this article and supplementary material.

## 7. Acknowledgements

This work was supported by JST Moonshot R&D Grant Number JPMJMS2011

# Stakeholder perspectives on designing socially acceptable social robots and robot avatars for Dubai and multicultural societies

## Supplementary Material

| Participant | Industry | Role |
| --- | --- | --- |
| P1 | Healthcare (hospital) | General director |
| P2 | Public transport (solutions development) | Operations director |
| P3 | Food & Beverage (service / delivery) | Product Lead |
| P4 | Retail - Hospitality Conglomerate | Head of Marketing |
| P5 | Airlines / Airport | Head of Robotics |
| P6 | Library | Business Analyst |
| P7 | Public transport | Manager |
| P8 | Food & Beverage (delivery) | Director Special Projects |
| P9 | Business Solutions provider with social robots | Chief Operations Officer |
| P10 | Retail | Training Manager, Operations, Customer care |

**Table 1A. Interviewee profiles**

**Questions asked during the interview***

PART I: Experience with social robot deployments

1. Do you have experience deploying robots and social robots at your organization to interact with your customers? Would you be able to share the most relevant use cases? What tasks did they perform? Where were the robots deployed, and for how long? How was the experience received among customers?

PART II: Robot avatars design insights

1. Can you identify a specific need or challenge that your customers face which a robot avatar could help address? For what spaces would they be suitable?

2. Thinking about your customer needs, what would be the ideal tasks for this robot avatar in your organization? (Tasks include providing information, customer registration, wayfinding, telepresence, patrolling, companionship, object delivery, transporting goods, survey data collection, and gathering feedback/complaints.)

3. We are going to show you different types of robot appearances. Imagine these were robot avatars teleoperated by a human. What level of acceptance would you expect among your customers for each type:

    - Ultra-realistic androids (e.g., robots with appearances very similar to a human being, with skin-looking texture).
    - Hybrid androids (i.e., robots that are a mix of human and robotic appearance).
    - Robotic-looking, highly anthropomorphic (i.e., robots that have body shapes and faces reminiscent of humans but clearly look robotic).
    - Robotic-looking, moderately anthropomorphic (i.e., robots that have minimal human-like features).
    - Cartoonish-looking robots.
    - Animal-looking robots.

4. What would be the ideal materials for the robot avatar's appearance? (Options provided: skin texture, plastic, metal, silicone.)

5. What should be the ideal height of the robot?

6. How many parts should the robot be able to move?

7. What shape should the robot have? (Options: rounded or squared.)

8. For limbs, what shape would be ideal? (Options: with hands and fingers or without.)

9. Should the robot be mobile or static?
10. Should the robot be physical or on a screen?
11. When the robot is autonomous, would you prefer the interactions to be verbal or on the screen?
12. When the robot is tele-operated, would you prefer the operator to appear on a screen or speak through the robot?
13. How should the robot express emotions? (Options include facial expressions, body movement, sounds, or emoticons.)
14. What color combinations should the robot have?
15. What level of acceptance would you expect for a robot avatar like the one you just described?

*Note. Given the semi-structured nature of the interview, the questions varied slightly depending on the characteristics of the organization